\documentclass[a4paper,11pt]{article}
\pdfoutput=1 

\usepackage[T1]{fontenc} 

\usepackage{bm}
\begin{document} 
\begin {center}{\Large	Nonlocal mass creation by metricized accelerations \\ in Euclidean matterspace}

\medskip
{Igor {\'E.} Bulyzhenkov}

\medskip
{Levich Institute for Time Nature Explorations, 
		Moscow,  Russia}
\end {center}






\abstract {Relativistic four-acceleration in metrically organized motions can introduce the nonlocal mechanism for the local generation of active and reactive mass densities in the monistic geometrodynamics of correlated matterspace. Four Hilbert variations for geometrized matterspace with Euclidean subgeometry maintain the universal equivalence of active and reactive mass densities. The vector counterbalance in the contracted Bianchi identities with the acceleration-based curvatures prevents the gravitational collapse of non-local self-organizations. 

\medskip
\textbf {Keywords} Classical Theories of Gravity, Space-Time Symmetries, Gauge Symmetry
}









\section{Introduction}
	\label{sec:intro}

\subsection {Monism versus Dualism}

In Europe, the concept of describing nature in terms of the properties of a material space-plenum or ponderable ether dates back to the worldview of the ancient Greeks.  \lq\lq Plato in the Timaeus says that matter and space are the same\rq\rq{} underlined Aristotle in his Physics~\cite {Ari}. Descartes\rq{s} matter-extension~\cite {Des,Des1}, from the seventeenth century, was supported  a hundred years later  by  Lomonosov with his ether-based gravity theory, which suggested local pushes by a material liquid with invisible densities~\cite {Lom}. 
 The pondarable space with inertial heat and etheric masses was first described quantitatively
by Umov\rq{s} continuous transport of energy flows in a monistic cosmos-medium~\cite {Umo}. This monistic distribution of continuous mass-energy densities can be reinterpreted and approximated by secondary (as Umov underlined) relations with volumetric masses in the
Newtonian empty space, but not vice versa.

  In Germany,  Gustav Mie persistently supported the field theory of matter since his first publication on the continuous elementary particle~\cite {Mie} in 1912, before the emerging era of quantum distributions in the non-local microworld. Einstein publicly accepted the ponderable ether in 1920~\cite {Ein1920}.
 In 1929, the New Yorker Magazine popularized Einstein by noting \lq\lq People slowly accustomed themselves to the idea that the physical states of space itself were the finite physical reality\rq\rq{.} For a long time, the monistic wave function for material densities of quantum particles was traditionally associated with the so-called microworld. The mechanical picture of the macroworld  relied on Newtonian dualism for localized bodies and continuous forces (or massless fields) in empty space. This dualistic consept dominated until non-locality of superconducting currents was established in the low-temperature laboratory. In this century, astronomers are publishing more and more about megaparsec correlations~\cite {Ali,Mys}.

\subsection {Did Einstein geometrize the void or the massive field?}

	In 1914, Einstein~\cite {Ein1914} proposed to nullify the geodesic  four-acceleration  $a_\mu \equiv u^\nu\nabla_\nu u_{\mu} \equiv u^\nu (\nabla_\nu u_{\mu} - \nabla_\mu u_{\nu})= u^\nu (\partial_\nu u_{\mu} - \partial_\mu u_{\nu})$ for the free fall of the probe body. By convention, the latter does not contribute to the external fields and the local derivatives $\nabla_\nu u_\mu \equiv \partial_\nu u_\mu -\Gamma_{\nu \mu}^\rho u_\rho$ with the symmetric affine connections (correlations) $\Gamma_{\nu \mu}^\rho = \Gamma_{\mu \nu}^\rho$.  In the twenties century, the revolution  4-acceleration $ a^\nu = g^{\mu\nu} a_\mu $ did not shake the particle-field dualism of matter. The point is that the relativistic mechanics and metric fields of Einstein, as well as the monistic energy transport of Umov,  were satisfactorily  referenced by the Newtonian formalism for many testable applications.

	In summer and autumn of 1915 Hilbert~\cite{Hil1915} counseltad Einstein regarding the
	dual action with the classical Lagrangian for inertial sources and the Ricci scalar
	density for metric fields (around these sources). As a result, the Einstein Equation for ten tensor balances between metric fields and massive densities~\cite {Ein}
	remained consistent with the Newtonian dualism for massive bodies and their
	forces or massless fields. In this dualistic approach to the monistic (quantum)
	reality, negative energies of gravitational or metric fields also remained as the
	independent notion, which ultimately led to the metric singularity and many
	mathematical surprises of post-Newtonian physics.

	For physical theories, practice is indeed crucial. But observations and measurements can only falsify a naive concept, not select the right one among competing theoretical constructions~\cite{Pop1959}. The dual partitioning of matter into massive particles and massless force fields in the Newtonian worldview leads to
	many computational absurdities, including the divergence of energy for a singularity. In 1939 Einstein completed the thought experiment~\cite{Ein1939} in order to reject
	Schwarzschild\rq{s} metric singularities for realistic kinds of mechanical movements
	- \lq\lq ... at least not faster than light\rq\rq{.}
	
	Why does the dense mass of probe bodies not affect the external field generated universally by all masses? Many dualistic shortcomings have encouraged
	researchers to reconsider sparse and dense energy regions in monistic terms of
	continuous mass. For example, \lq\lq  A thrown stone is, from this point of view, 
	changing field, where the states of greatest field intensity travel through space
	with the velocity of the stone\rq\rq{} Einstein and Infeld whrote about the non-dual
	(monistic) reinterpretation of the observable reality~\cite{EI1938}. We tend to maintain
	the relativistic four-vector $a^\mu$ as a primary notion in the mass-generation mechanism. But how to build auto-dynamics of inertial spaces with locally induced
	masses by metrically correlated accelerations in order to quantitatively describe
	nonlocal self-organization of mechanical macrodistributions with negligible dissipation losses?

	\subsection { Macroscopic entanglement revives material space-plenum}

Despite the original intention of Hilbert to combine  Mie\rq{s} theory of field matter~\cite {Mie} with Einstein\rq{s} relativistic ideas~\cite {Ein1914}, the monistic macromechanics of inertial, massive fields was not the mainstream of theoretical studies during the last century.  However, the 2022 Nobel Prize awarded in Physics for the nonlocal macro-experiments encouraged theorists to study long-range entanglement in the correlated macrocosm. Nonlocality in any closed hierarchical subsystem should survive until dissipative intervention or measurement, as in the mechanical microcosm of quantum nonlocality.

 Nowadays the Cartesian mechanics of inertial space is experiencing a renaissance due to new experimental data, including astro-correlations~\cite {Ali,Mys}. This paper presents covariant updates for the static equilibrium of extended mass-energy, where the Ricci scalar density $R(r) \equiv (
 \mu_a + \mu_p) 8\pi G/c^2 $ $= 4 r_o^2/[r^2(r+r_o)^2]$ is proportional to the equal densities of active and passive (or reactive) masses~\cite {Bul2008}. The monistic macrophysics of  inertial fields or field masses aims to clarify how inhomogeneous densities of kinetic energy resist gravitational collapse or resolve the Bentley paradox of Newtonian physics.  The goal of our
 flatspace relativism, with only four Hilbert variations under the six geometric
 constraints or inherent symmetries of space-time geometry, is to maintain Cartesian worldview with its matter-extension quantitatively. Additionally, we aim
 to initiate the adaptive auto-dynamics of metrically correlated inertial densities
 (called below geometrodynamics) for further applications in nonlocal macrophysics, life sciences, engineering, and the novel energo-informatics of big data.

\section{Accelerations define field masses and monistic action }

\subsection
{Pseudo-Riemann definitions of material space with its
local velocities and inertial currents}

According to the space-time geometrization of static distributions with spherical symmetry, the universal self-potential $\varphi_{\!_G} \equiv c^2/{\sqrt G}$ of the isolated mechanical charge
${\sqrt G} m \equiv r_o c^2/{\sqrt G} = const$ provides the rest energy integral ${\sqrt G}m\varphi_{\!_G}  \equiv mc^2 = \int dV^\prime \mu ({x^\prime}) c^2 = const$ with the radial  mass density $ \mu(r^\prime) = mr_o/[4\pi {r^\prime}^2(r^\prime+r_o)^2] \Rightarrow \mu(x,y,z) \sqrt {1-\beta^2}  $ in the co-moving local frame $\{x^{\prime o}/c \equiv t^\prime; x^{\prime i}\}$~\cite {Bul2008}.
 Similarly to Maxwell\rq{s} physics of electric currents, geometrodynamics can relate the mechanical current density $j^\nu(x) \equiv c \varphi_{\!\!_{_G}}\nabla_\mu f^{\nu\mu}(x) / {4\pi } \equiv c \varphi_{\!\!_{_G}}i^\nu (x)/ {4\pi }$ of material space (called matterspace) to the antisymmetric tensor $f^{\mu\nu}(x)\equiv \nabla^\mu u^\nu(x) -\nabla^\nu u^\mu(x) $ with the continuous field of local four-velocities $cu^\mu (x)\equiv c dx^\mu / {\sqrt {g_{\rho\lambda} dx^\rho dx^\lambda}} $ $\equiv cg^{\mu\nu}u_\nu$:
\begin {eqnarray} \label {eq:1}
\cases{ 
	u_\mu u^\mu \equiv 1, \	f^{\mu\nu} (x)
	\equiv
	g^{\mu\lambda}(x) g^{\nu \rho}(x) f_{\lambda \rho} (x) \equiv - 	f^{\nu\mu} (x), 
	\cr
	f_{\mu\nu}(x) \equiv  
	\nabla_\mu u_\nu(x) - \nabla_\nu u_\mu(x)
	= \partial_\mu u_\nu(x)
	- \partial_\nu u_\mu (x), 
	\     
	\cr 
	\nabla_{\mu} f_{\nu\lambda}(x) + \nabla_{\nu} f_{\lambda\mu}(x)+\nabla_{\lambda} f_{\mu\nu}(x) \equiv 0,  
	\cr  
	j^\mu(x)\!
	\equiv \!  {c\varphi_{\!\!_{_G}}\partial_\nu \!\left [{\sqrt {-g(x)}}  f^{\mu\nu}(x)\right ]  } / {4\pi \sqrt {-g(x)}},  \ \nabla_\mu  j^\mu(x) \equiv 0, 
	\cr
	a^\mu(x)  \equiv  u_\nu f^{\nu\mu}(x) , \  u^\nu \nabla_\mu u_\nu  \equiv  u_\nu \nabla_\mu u^\nu \equiv 0,  \    u_\mu(x) a^\mu(x) \equiv \! 0 .
}
\end {eqnarray}

The static self-organization of the nonlocal charge in the laboratory achieves equilibrium energy density $ \varphi_{\!_G}{\sqrt G}\mu (r^\prime = r) \equiv c^2\mu (r^\prime = r) $ around the point vertex at the coordinate origin. 
This static distribution follows the Poisson equation $ \mu(r) = 
 -  \delta^{ij} \partial_i w_j/4\pi G  $ 
with the metric field intensity $w_i(r) = \delta_{ij}w^j=  - Gm (x^j/r) /[r(r+r_o)] \equiv - c^2{\partial_i}    ln {\sqrt {g_{oo}(r)}} $, $g_{oo} (r)= r^2/(r+r_o)^2 = 1/ g^{oo}(r) $~\cite {Bul2008}. However,  inhomogeneous profiles of extended masses and their rest energies require stabilization by non-Newtonian forces, such as those related to the so-called Poincar{\'e} pressure~\cite {Poi}. How can we balance the self-organizing forces in the monistic geometrodynamics of material space with affine connections (used in mathematics) or nonlocal correlations (used in physics)?

\subsection {Monistic action for nonlocal masses with metrically
	correlated four-currents}

The essential message of the radial solutions in~\cite {Bul2008} for static inertial  distributions  was the equivalence  of the active (attractive) and passive (reactive)  mass-energy densities, $\mu_a(r) c^2 \equiv -  c^2 \partial_i w_j (r) \delta^{ij} /4\pi G $ $  = w_i (r)w_j(r)  \delta^{ij} /4\pi G \equiv \mu_p (r)c^2$. Both densities contribute to the Ricci scalar $R (r) = (\mu_a c^2+ \mu_p c^2) 8\pi /\varphi_{\!_G}^2 $ $  = 4 r_o^2/[r(r+r_o)]^2 = 2 u_o i^o(r) - 2f_{oi}(r)f^{oi}(r) $ in the co-moving frame with $u_o = {\sqrt{ g_{oo}(r)}} \equiv g_o(r) = r/(r+r_o), g_i \equiv g_{oi}/ g_o = 0,  g_{oi} = 0, g^{oi} = 0, g^{ij} = -\delta^{ij} $, $g_{ij} = - \delta_{ij}$, $u_i = u^i = 0$, $f_{oi} = - \partial_i g_o =  - (x_i r_o /r) / (r+r_o)^2, f^{oi} = g^{oo}g^{ij}f_{oj} = \delta^{ij} x_i r_o/r^3, (-f_{oi}f^{oi}) =  r^2_o /[r(r+r_o)]^2 = g_o i^o, i^o = r_o^2/[r^3(r+r_o)]$.  The invariant sum of equal static contributions clarifies the tensor structure of the Lagrangian ${\cal L} (x)= -\varphi_{\!_{_G}}^2 B(x)/ {16\pi} $ with the scalar density   $B(x)  = 2u_\nu (x)i^\nu(x) - f_{\mu\nu}(x)f^{\mu\nu}(x) > 0$ in the  monistic action $S_m$ for inertial (massive) fields by the local divergence of the coordinate self-accelerations $a^\mu(x) \equiv u_\nu(x) f^{\nu\mu}(x)$:   
\begin {eqnarray}\label{eq:2}
S_m \equiv   
- \!\int\!  \frac {{ \varphi_{\!_{_G}}^2}\!d^4x } {16\pi c} { \sqrt { -g}}     \left ( \!2u_\nu \!\nabla_\mu f^{\nu\mu}\!-\! f_{\mu\nu}\! f^{\mu\nu}\! \right )   
\!\equiv \! 
-\! \int\!  \frac {{ \varphi_{\!_{_G}}^2}d^4x } {8\pi c}  \sqrt { -g} 
\nabla_\mu a^\mu 
\cr \equiv
- \int\! \frac {{ c^3}d^4x } {8\pi G}  \sqrt { -g} [\nabla_o a^o (x)+ \nabla_i a^i(x)]\equiv -
\frac {c } {2 } \int\! 
{{\sqrt { -g}}d^4x }   [\mu_p(x)  + \mu_a (x)] .     
\end {eqnarray} 

By defining the monistic field tensor $T_{\mu\nu}(x)$
from the Hilbert formula for the substance stress-energy tensor~\cite {Hil1915} ${\sqrt {-g}}T_{\mu\nu}/2 \equiv \partial ({\sqrt {-g}} {\cal L})/\partial g^{\mu\nu}$ with the opposite sigh (to match the dualistic balance of Hilbert fields and gravitating substances)  and using $\partial {\sqrt {-g}} \equiv -{\sqrt {-g}}g_{\mu\nu}\partial g^{\mu\nu} /2 $ for the metric space-time, one gets the following notations for inertial densities of continuous matter in monistic field theory:
\begin {eqnarray}\label{eq:3}
\cases{ 
	\frac {{\sqrt {-g}}}{2}T_{\mu\nu} \equiv  
	\frac {c^4{\sqrt { -g}} }{16\pi G} (g_{\mu\mu} \nabla_\lambda a^\lambda -  \nabla_\mu a_\nu  - \nabla_\nu a_\mu   ), \cr
	T^\nu_\mu \equiv g^{\nu\lambda}T_{\mu\lambda} \equiv 
	\frac {\varphi_{\!_{_G}}^2}{8\pi} (\delta_\mu^\nu  \frac {{ B}}{2} -  {B}_\mu^\nu     ),  \ B_{\mu\nu} \equiv
	\nabla_\mu a_\nu  + \nabla_\nu a_\mu,   
	\cr {B}^\nu_{\mu} \equiv 
	g^{\nu\lambda}B_{\mu\lambda},   B^\mu_{\mu} \equiv 2 \nabla_\mu a^\mu  \Rightarrow \frac {4r_o^2}{r^2(r+r_o)^2},  a^o = 0, u^i = 0,
	\cr
	T^\mu_\mu \equiv \frac {\varphi_{\!_{_G}}^2}{4\pi }\nabla_\mu a^\mu \equiv
	\frac {\varphi_{\!_{_G}}^2}{4\pi }\nabla_o a^o \! + \!  \frac {\varphi_{\!_{_G}}^2}{4\pi }\nabla_i a^i   \equiv (\mu_p\!+\!\mu_a)c^2 \equiv \frac {\varphi_{\!_{_G}}^2}{8\pi} 
	( B_o^o\! +\! B^i_i),
	\cr 
	\mu_p c^2 \equiv \frac {\varphi_{\!_{_G}}^2}{4\pi }\nabla_o a^o  \equiv \frac {\varphi_{\!_{_G}}^2}{4\pi }(\partial_o a^o + \Gamma_{o\lambda}^o a^\lambda)\Rightarrow \frac {\varphi_{\!_{_G}}^2}{4\pi } \Gamma_{oi}^o a^i = 
	\frac {\varphi_{\!_{_G}}^2r_o^2}{4\pi r^2 (r+r_o)^2},
	\cr \mu_a c^2 \equiv \frac {\varphi_{\!_{_G}}^2}{4\pi }\nabla_i a^i  \equiv \frac {\varphi_{\!_{_G}}^2}{4\pi }(\partial_i a^i + \Gamma_{i\lambda}^i a^\lambda)\Rightarrow  \frac {\varphi_{\!_{_G}}^2}{4\pi }\partial_i a^i =  \frac {\varphi_{\!_{_G}}^2r_o^2}{4\pi r^2 (r+r_o)^2}.
}
\end {eqnarray}
The Einstein-type tensor $g_{\mu\nu}(B^\lambda_\lambda/2) - B_{\mu\nu} \equiv 8\pi T_{\mu\nu} /{\varphi_{\!_{_G}}^2} $  and its tensor curvature  $B_{\mu\nu} \equiv  \nabla_\mu a_\nu + \nabla_\nu a_\mu$ $= R_{\mu\nu} + \theta_{\mu\nu} $, where $R_{\mu\nu}  \equiv \partial_\lambda \Gamma^\lambda_{\mu\nu} -\partial_\nu \Gamma^\lambda_{\mu\lambda} + \Gamma^\lambda_{\mu\nu}\Gamma^\rho_{\rho\lambda} -\Gamma^\lambda_{\mu\rho}\Gamma^\rho_{\nu\lambda}  $,  $ g^{\mu\nu} \theta_{\mu\nu} \equiv 0$,  $g^{\mu\nu}B_{\mu\nu}  =g^{\mu\nu}R_{\mu\nu}\equiv R  $ and $B_{\mu\nu}\neq R_{\mu\nu}$, are based  in our novel  analyses on the covariant derivatives of four-accelerations  $a_\mu$. Hereinafter we require strict 3D flatness, $\kappa_{ij} \equiv g_ig_j - g_{ij} \equiv \delta_{ij}$, for the monistic continuum of field masses. The corresponding six constraints for Euclidean 3-space, usually required by theorists for electrodynamics and quantum mechanics, have not yet been falsified in the available experiments. The metric component $g_{oo} = r^2/ (r+r_o)^2$ for the radial self-organisation of static material continuum~\cite {Bul2008} is free from pecularities. Such metric field allows to recalculate the main predictions of the dualistic General Relativity (gravitational light bending, Shapiro ekho delay, Merciry perehelion precession) in terms of flat space with adaptively delayed local time.

The stress-energy tensor $T_{\mu\nu}(x) \equiv g_{\nu\lambda}(x)T^\lambda_\mu (x)$ elucides the local generation of non-local (field) masses, highlighting the non-Newtonian monism of material reality and invariant fields pivotal in modern metric theories. Their geometric interplay bridges classical and quantum physics, inspiring cosmological models and advancing geometrodynamic paradigms. 

\subsection {Inherent space-time symmetris for Euclidean 3-geometry}

The Euclidean constraints $\kappa_{ij} = \delta_{ij}, g^{ij} = \delta^{ij}$  can be called the inherent
symmetries of the space-time geometry in the basic geometrodynamic equality $u_\mu u^\mu \equiv 1$.
These symmetries are maintained by the adaptive local time $d\tau \equiv g_\mu dx^\mu/c  \neq dx^o/c$, 
 where $g_\mu \equiv g_{o\mu}/g_o$. Besides beying a function of metric
and motion changes, the local time for physical processes is not a primary concept
in the correlated field theory for a closed metric (hierachial) system. It is a
secondary variable that depends on energy-momentum variations, but not vice
versa. However, the common shift $dx^o/c$ of the universal (world, Newtonian) parameter $x^o/c$
can
be considered as a primary notion, like three Euclidian (universal) coordinates $x^i$ for the Cartesian matter-extension. Essentially, this implies that four temporal
components $g_{o\mu} $ together with the metric 4-potential  $g_\mu c^2$ and the corresponding time rate 
(experienced locally by a probe body) are consistent with  observable consequences. Time and metric variations are not a primary cause of observable phenomena, but a nonlocal consequence of the correlated kinetics of all massive fields.

For static self-organisation of massive fields with spherical symmetry, where $ \gamma \equiv (1-\beta^2)^{-1/2} = 1, u_o = {\sqrt{ g_{oo}(r)}} \equiv g_o(r) = r/(r+r_o), g_i \equiv g_{oi}/ g_o = 0,  g_{oi} = 0, g^{oi} = 0, g^{ij} = -\delta^{ij} $, $g_{ij} = - \delta_{ij}$, $u_i = u^i = 0$,
one can use the already known metric solutions:
 $f_{oi} = - \partial_i g_o =  - (x_i r_o /r) / (r+r_o)^2, f^{oi} = g^{oo}g^{ij}f_{oj} = \delta^{ij} x_i r_o/r^3, (-f_{oi}f^{oi}) =  r^2_o /[r(r+r_o)]^2 = g_o i^o, i^o = r_o^2/[r^3(r+r_o)]$.  These metric fields with radial inertial densities correspond to $a^o \equiv u_i f^{oi} =0, a_o \equiv u^if_{io} = 0$,  $a^i = u_o f^{oi} 
 \equiv u_o g^{o\lambda}g^{\rho i}f_{\lambda\rho} =
 u_o g^{oo}g^{i\rho}f_{o\rho} = - g_o g^{oo}f_{oj}\delta^{ij}
 = \delta^{ij}\partial_j ln g_o = \delta^{ij} \Gamma^o_{oj} = x_i r_o/[r^2 (r+r_o)]$, and $a_i = u^of_{oi} = - \partial_i ln u_o = - \Gamma^o_{oi} = - \delta_{ij} a^j$.
 Thus,
 the 2007 solution~\cite{Bul2008} for the strong-field interactions with probe (external)
 masses, $w_i (r)\equiv - Gm \delta_{ij} x^j / [r^2 (r+r_o)] = a_i (r) c^2$, is determined by the Einsteinian
 3-acceleration $c^2a_i(r)$ of the static material densities within their correlated distribution in the volumetric mass integral $m$.

 \subsection {Primacy of local accelerations for field matter observations by probe bodies}

The relativistic four-acceleration $a_\mu \equiv g_{\mu\nu}a^\nu$ is a primary characteristic of the local creation and dynamic self-distribution of massive densities in the  nonlocal mass-energy  $mc^2 = const$. The action (2) 
with $\nabla_\mu a^\mu (x)$ would be dropped in the dualistic mechanics for localized masses near the coordinate origin due to the Gauss theorem for the distant hypersurface. 
The nonlocal mass generation
mechanism associated with $\nabla_\mu a^\mu (x)> 0$
 in the monistic action of massive fields
has no analogues in dualastic Lagrangians of localized masses and massless fields on distant surfaces.

In the monistic mechanics for quasi-isolated (geometrized) distributions both
geometric fields and inertial currents are defined by the local self-acceleration  $a^\mu$ of matterspace with adaptive local times. The 3-acceleration $a^i \equiv g^{i\nu} a_\nu \equiv  u_\nu f^{\nu i} \neq 0 $ is  responsible for everything in static self-organizations ($g^{oi}= 0, g^{oo} = 1/g_{oo} > 1, a^o = 0$,  $a^i = -\delta^{ij} a_j$),  including the generation of both active, $\mu_a = c^2\partial_i a^i /4\pi G $, and passive, $\mu_p = c^2a^i (\partial_i ln g_o) /4\pi G $, mass densities or Maxwell-type currents of mechanical energy.The spatial 3-divergence is well known as a
local density of (active) charges in Coulomb forces. The absence of matterspace
accelerations relative to the local observer implies that neither the passive mass
density nor the active mass density cannot emerge in the geodesically falling
elevator for the practical detection of dense materspace for non-geodesically
moving observers.

In mechanics, the acceleration three-vector $a_i = -\delta_{ij}a^j$ works as the interaction field for the probe mass. The temporal divergence $\nabla_o a^o =  \Gamma_{oi}^o a^i = -a_i a^i > 0$ seems to be a local self-action associated with the metric time auto-changes $cd\tau/dx^o =  {\sqrt {g_{oo}}}  = r/(r+r_o) \neq const $. The Minkowski metric with $g_{oo} = 1, g_{ij} = -\delta_{ij}$ has no local self-accelerations, $ a_o=a^o = a_i = a^i = 0$, $\nabla_o a^o = \nabla_i a^i  = 0$, and can produce neither the passive mass density $\mu_p$, nor the active $\mu_a$. Following the original initiative of Grossmann in 1912, the co-authors made only temporal changes in the Minkowski 4-interval, $ds = \sqrt {c^2 d\tau^2(g_{\mu}) - \delta_{ij}dx^idx^j}$, preserving  Euclidean 3-space~\cite {GE1912}. 
 Again, metric time dilation can quantitatively explain all post-Newtonian corrections (light bending, Shapiro echo delay, perihelion precession, etc.) and to propose the acceleration mechanism for mass generation. It now appears that Grossmann was reasonably reluctant to
 endorse or sign Einstein\rq{s} later manuscripts with the hypothetical curvature of
 3-space in the advised 4D pseudo-Riemann geometry.

\section {Non-local geometrodynamics of Euclidean matterspace
with adaptive time rate}

\subsection {Equivalence of mass densities from four variational
	equations}

In a flat matterspace characterized by affine connections over the whole volume, the nonlocal self-organization of correlated densities in the closed system
always satisfies the basic conditions: $u_\mu u^\mu \equiv 1$  and $u_\mu a^\mu \equiv 0$. The specific 4D
geometry of each closed (elementary) system preserves local properties without
distorting the universal 3D Euclidean geometry for the hierarchy of superimposed 3-volumes of different geometric systems. The requirement to preserve
the universal subgeometry of 3D physical spaces for the multi-system universe
with nonlocal correlations leads to the reduction of the ten variables in the
Hilbert formalism to four. The effective geometrodynamics of each metric system, in particular with respect to its mass-energy conservation, is based on the
correlation of the specific metric potentials $ g_\mu c^2 $ and 
the specific adaptivity of the local time rates  
$cd\tau/dx^o \neq 1$.

Again, time-varying and equilibrium self-organizations of correlated densities
in nonlocal matterspace maintains everywhere  the curved 4D geometry
of 3+1 manifolds with adaptive time rates for the  flat spatial continuum. Such self-consistent geometry never distorts the Euclidean 3D subgeometry in adaptive
self-movements within the nonlocal unity. The theoretical requirement to preserve Euclidean 3-space by six inherent symmetries $g^{ij} \equiv - \delta^{ij}$ or $g_{ij} \equiv g_ig_j - \delta_{ij}$
reduces ten Hilbert variables $g^{\mu\nu} = g^{\nu\mu}$~\cite {Hil1915} to four,  $ g^{\mu o} = g^{o\mu}$.
 One can eliminate the spatial variations in the Hilbert formalism, $T_{ij}\delta g^{ij} \equiv T_{ij}\delta (-\delta^{ij}) = 0$ in spite of $T_{ij}(x) \neq 0$.

 Adaptive correlations in the geometrodynamics of continuous matterspace are maintained only by relevant responses of four metric potentials $g_\mu \equiv g_{oi}/g_o$ and the local time rate $ d\tau \equiv g_\mu dx^\mu/ c$, with 
  $c^2 d\tau^2 \equiv g_{\mu\nu} dx^\mu dx^\nu + \delta_{ij} dx^idx^j $, for all local changes within the constant integrals of volumetric mass-energy, 3-momentum, and angular momentum.  There are no geometric restrictions for the time related variations,   
$\delta g^{\mu o} = \delta g^{o\mu}\neq 0$, in the flat-space formalism. Here one can expect only four independent equations after the metric variations,  but not ten tensor balances as in the Einstein Equation for dualistic gravitation in the warped 3D void around Newtonian masses~\cite {Ein}. Four variational equations of continuous matterspace declare $T_{\mu o} = T_{o\mu} = 0$ and leads quantitatively to the local equivalence of active and passive mass densities in all static and dynamic relations: 
\begin {eqnarray} \label{eq:431}
\cases{
	T_{\mu o}  \equiv\frac {\varphi_{\!_{_G}}^2}{8\pi} (g_{\mu o} \nabla_\lambda a^\lambda -  \nabla_\mu a_o - \nabla_o a_\mu  ) 
	\equiv \frac {\varphi_{\!_{_G}}^2}{16\pi} (g_{\mu o}B - 2B_{\mu o} ) = 0, 
	\cr g_{oo} {B} = 2B_{oo} = 4\partial_o a_o - 4\Gamma_{oo}^\lambda a_\lambda, 
	\  g_{oi} {B} = 2B_{oi} = 2(\partial_o a_i + \partial_i a_o)  - 4\Gamma_{oi}^\lambda a_\lambda, 
		\cr
		T^\nu_{o } \equiv \frac {\varphi_{\!_{_G}}^2}{8\pi} (\delta^\nu_o \nabla_\lambda a^\lambda -  \nabla_o a^\nu  - \nabla^\nu a_o )	\equiv \frac {\varphi_{\!_{_G}}^2}{16\pi} (\delta_o^\nu B - 2B_{o}^\nu ) = 0,  \cr  B_o^i  \equiv g^{i\lambda}B_{o\lambda} \equiv  \nabla_o a^i + g^{o i} \nabla_o a_o + g^{j i} \nabla_j a_o = 0,   
		\cr  	\nabla_\lambda a^\lambda \equiv \nabla_o a^o + \nabla_i a^i  = 2  \nabla_o a^o =
	 2  \nabla_o a_o/g_{oo}   ,\cr
		\nabla_i a^i \equiv  B^i_i /2 =   B^o_o /2 \equiv  \nabla_o a^o\ $or$ \ \mu_a(x) = \mu_p(x)
		\cr		
				 T^o_o \equiv \frac {\varphi_{\!_{_G}}^2}{16\pi} (B^i_i-B^o_o) = 0, \ B^o_o = B^i_i =  \frac {1}{2} B  \equiv \frac {1}{2} (B^o_o + B^i_i ) > 0,
		\cr
	 T^i_i =	T^\mu_\mu  \equiv \frac {\varphi_{\!_{_G}}^2}{8\pi}  B^\mu_\mu =  \frac {\varphi_{\!_{_G}}^2}{4\pi}  B^i_i \equiv 2\mu_a c^2 = 2\mu_p c^2  \equiv \frac {\varphi_{\!_{_G}}^2}{4\pi}  B^o_o .
	}
\end {eqnarray}

The trace of the energy-momentum tensor in the monistic field theory is
equal not to zero, but to the positive sum of two equal densities: 
$\mu_a (x)c^2 + \mu_p (x) c^2 = T^\mu_\mu (x) > 0. $
 The time-varying mass equality, $\mu_a(x^i, x^o) = \mu_p(x^i, x^o)$ for the active density  $\mu_a (x)\equiv $ ${\varphi_{\!_{_G}}^2}\nabla_i a^i(x) /{4\pi } $ $\equiv c^2 B^i_i(x) /8\pi G$  and the passive one $\mu_p(x) \equiv {\varphi_{\!_{_G}}^2}\nabla_o a^o (x)/{4\pi }$ $ \equiv c^2 B^o_o(x) /8\pi G $,  was derived in (4) from the Hilbert
variations as the Lagrange equation of motion, but not postulated from the
General Theory of Relativity as a principle. The positive scalar densities continuously fill the correlated 3-volume with the fixed mass-energy integral,  $const = $ $\int d^3 x   {\sqrt {-g}} [
\mu_a  (x^i, x^o)  + \mu_p (x^i, x^o)]$.

The spatial density $B (x^i, x^o)> 0$ is the scalar curvature for the 4D self-organization. This density provides the positive mass-energy integral to the observable 3D world with the universal 3D subgeometry. The framework presented reflects
an inextricable link between active and reactive mass densities, based on geometric principles that maintain the integrity of physical laws across different
frames of reference. The quantitative equality of active and reactive mass densities, which are different functions of the relativistic acceleration, underlines
the essential nature of metric transformations and the correlated geometrodynamics. Monistic matterspace with universal 3D geometry and adaptive time
unifies field masses and thier energy densities in a nonlocal material volume of
the pseudo-Riemannian 4D organization.

 \subsection {Zero energy integral for a closed mechanical self-assembly}
  
  The adaptive 4D manifolds with specific times in different auto-organizations of
  quasi-isolated systems maintain zero variational densities $T_{\mu o} (x)= 0$ resulting in $T^\nu_o (x) = g^{\nu\mu}(x) T_{\mu o}(x) =0$, while $T^{\mu o}(x) \equiv  g^{o\nu} T^\mu_\nu = g^{oi}T_i^\mu \neq 0$. The system mechanical
  energy for nonlocally correlated material densities vanishes after 3D integration
  over the hypersurfaces, 
  \begin {equation} \label {eq:532}
\int d^3 x {\sqrt {-g}} T_{o\mu}(x)d\sigma^\mu = 0.
  \end {equation}
  This finding from the Hilbert variations with Euclidean 3D section seems strange
  for inertial fields with only positive mass densities. However, the zero energy integral of a closed mechanical system with self-gravity was intuitively discussed
  (by P. Dirac, P. Jordan, R.J. Glauber and others) in the 1930s-1950s period. Dualistic physics exploited the idea of negative binding energy in isolated systems
  like a secluded star. But General Relativity with warped 3-space failed in quantitative calculations of strong field energy balances for the sum of interacting
  masses. The flat-space result (5) 
  corresponding to the equivalence of active and
  passive mass densities, is deciphered in the last section by analytical examples
  with the Shannon information within a continuous energy distribution.

  The massive field self-organization with $T^\nu_o (x^i, x^o)= 0 $ and $ T_i^\nu (x^i, x^o) \neq  0$  means that there is no 4D energy continuum for observations. Only 3D space $x^i$ is continuously filled by the measurable mass-energy, while $x^o$ is not the continuum coordinate. Rather, it is a world running parameter to form a specific metric time for each quasi-isolated system within the common spatial superposition under   universal 3D geometry.  If the external forces are neglected, the quasi-closed system is described by metric-kinetic flows of mass-energy in flat space. Both the corresponding components of the symmetric metric tensor $g_{\mu\nu}$ and the affine connections $\Gamma^\lambda_{\mu\nu}$ can be expressed in terms of 
 $g_\mu \equiv g_{o\mu} / {\sqrt {g_{oo}}}$ or   
  the metric 4-potential $\varphi_{\!_{_G} } g_\mu$ when  3D continuum of field masses is flat:   
 \begin {eqnarray}\label{eq:633}
 \cases{ 
g_{ i j} = g_i g_j - \delta_{i j}, \ g^{ij} = - \delta^{ij}, \  \delta^{ij} g_i =  g_og^{oj} \equiv g^i, \ g^{oo}= (1-g_ig_j \delta^{ij})/g_o^2, 
\cr \Gamma^\lambda_{\mu\nu} \equiv \frac {g^{\lambda\rho}}{2} (\partial_\mu g_{\nu \rho} + \partial_\nu g_{\mu \rho} - \partial_\rho g_{\mu \nu}  ), 
\cr
\Gamma^o_{oo} =  \partial_o ln g_o + \delta^{ij} g_i (\partial_o g_j- \partial_j g_o), 
\ \Gamma^i_{oo} =   \delta^{ij} g_o (\partial_j g_o - \partial_o g_j),
\cr  \Gamma^o_{oi} =  (1-g_kg_j \delta^{kj}) \partial_i ln g_o - 
\frac { \delta^{kj}g_k}{2g_o} [  \partial_o (g_j g_i) + \partial_i (g_o g_j)  - \partial_j (g_o g_i)  ],
 \cr  \Gamma^i_{oj} = \delta^{ik} g_k  \partial_j g_o - 
 \frac {\delta^{ik}}{2} [  \partial_o (g_j g_k) + \partial_j (g_o g_k)  - \partial_k (g_o g_j)  ], \cr
 \Gamma^k_{ij}\! =\! \frac {\delta^{kl}}{2}\! \left [\frac {g_l}{g_o}[\partial_i  (g_o g_j)\! + \!\partial_j
 (g_o g_i)\! - \! \partial_o (g_i g_j)  ]\!
 + \!\partial_l   (g_i g_j)\! -\! \partial_i
 (g_l g_j)\! - \! \partial_j (g_l g_i)  \right ],
 \cr
  \Gamma^\nu_{\mu\nu} =  \frac {g^{\nu\rho}}{2} \partial_\mu g_{\nu \rho} = (1-g_ig^i)\partial_\mu ln g_o + \frac {g^i}{g_o}\partial_\mu (g_og_i) - \frac {\delta^{ij}}{2}\partial_\mu (g_ig_j) = \partial_\mu ln g_o .
	}
\end {eqnarray}

  These expressions form the basis for determining how the metric behaves under transformations that satisfy the requirements of tensor covariance. Not arbitrary, but physically admissible transformations between 3+1 manifolds with
  inherent metric symmetries for Euclidean subgeometry, where
  $|g_{ij}(x)|$ $ = 1$ and
 $ {\sqrt {{-g}(x)}} = g_o(x)$, are responsible for the covariance in equations (4).
 The adaptive dilation 
 $cd\tau (x)/dx^o \neq 1$
 of the proper (physical) time rate  in
  the geometric organization is not a primary entity to define the material processes, but a secondary function or observable phenomenon in the world picture
  and its description. Again, the physical time does not act as a primary
  driver of material processes. Rather, it emerges from the metric organization of
  the specific 4D manifold, acting as a secondary consequence of its geometrodynamics.

 \subsection {Geometrodynamic counterbalance in non-equilibrium self-organisations}

The covariant divergence of $T^\nu_\mu (x)$ should vanish in the Hilbert formalism due to the Noether theorem. The geometrized dynamics of the correlated continuum of inertial fields obeys the contracted Bianchi identity  with the curvature tensor defined as  ${B}_{\mu\nu} (x) \equiv \nabla_\mu a_\nu  + \nabla_\nu a_\mu$,
 $B_\mu^\nu \equiv g^{\nu\lambda} B_{\mu\lambda}$, and $B \equiv 2\nabla_\mu a^\nu$:
\begin {equation}\label {eq:734}
\nabla_\nu T^\nu_\mu \equiv  \frac {\varphi_{\!_{_G}}^2}{8\pi} \nabla_\nu (\delta_\mu^\nu \frac {B}{2} - B_\mu^\nu     ) 
	\equiv \frac {c^4}{8\pi G} (\nabla_\mu \nabla_\nu a^\nu -  \nabla_\nu \nabla_\mu a^\nu    - \nabla_\nu \nabla^\nu a_\mu  ) = 0.
	\end {equation}
This covariant consequence of correlated stresses in any pseudo-Riemannian space-time reads for local  4-accelerations $a_\mu(x)$ of emerging mass densities as the  geometric feedback
$ \nabla_\nu \nabla^\nu a_\mu = -  R_{\mu\nu} a^\nu$.

One can study geometrodynamic balances along 3 spatial axes 
$x^i$ in the material continuum by rewriting the tensor contraction (7) 
in another equivalent form using 3D geometric
constraints  ${\sqrt {-g(x)}} = g_o(x)$ and the variational consequence $T^\nu_o (x)$ $= 0$ for non-local self-assembly, where $T^\nu_i (x) \equiv g^{\nu \mu}T_{i\mu} (x)= g^{\nu j}T_{ij} (x) $ $\neq 1$ and  $g_oT^{oo} \equiv g_og^{\nu o}T_{\nu}^o  (x)= g^{j}T^o_{j} (x) $ $\neq 1$:
\begin {eqnarray}\label {eq:835}
\nabla_\nu T^\nu_i = \frac {\partial_\nu ({\sqrt {-g}} T^\nu_i) }{\sqrt {-g}}- \frac { T^{\mu\nu}} {2}\partial_i g_{\mu\nu} =  
 \frac {\partial_\nu ({g_o} g^{\nu j}T_{ij}) }{g_o} 
 \cr
 -\delta^{kj}g_k T^o_j \partial_i g_o  - T^{oj}\partial_i (g_og_j)
   - \frac {T^{kj}} {2}\partial_i (g_{k}g_j) = 0.
\end {eqnarray}

The interplay of the derivatives with four metric potentials suggests how
the geometry is embedded in the adaptive geometrodynamics of the emerging
mass densities in the energy-momentum tensor. The covariant counterbalance
in (7) or (8) 
for non-equilibrium oscillations of elementary matterspace around
the equilibrium profile of inertial densities implies that the geometrodynamic
self-assembly maintains stability under perturbations. The tensor expressions
for the 4-velocities and 4-accelerations of the generated mass densities show
how these coordinate fields are intertwined with the emerging 4D geometry of
a specific hierarchical system.

Inhomogeneous mass distributions cannot exist in static states in Newtonian
dual mechanics and gravitation. This dualistic theory was developed for energy
and momentum exchanges between separated partners in a void, rather than
for nonlocally controlled self-accelerations of the elastic whole as a nonlocal
continuum in quantum mechanics.
The radial mass-energy densities of the static matterspace, as described in reference~\cite {Bul2008}, are given by:
\begin {eqnarray}\label {eq:936}
\mu_p (r) c^2 =  \mu_a (r) c^2 =  \frac  {c^4 B(r)}{16\pi G} = \frac {Gm^2 c^4} {4\pi r^2 (c^2r+Gm)^2} \Rightarrow \frac  {c^4 }{4\pi G r^2}. 
\end {eqnarray}
This continual solution shows that there is universal curvature near the coordinate  origin,  
$lim_{r\rightarrow 0} R(r \ll Gm/c^2)$ $ = {4}/{ r^2}$. For relatively large distances from the center of spherical symmetry, $r \gg Gm/c^2$,  the radial mass densities are
proportional to $m^2/r^4$, which is the self-action of Newtonian interaction fields.

In the monistic physics of the nonlocal whole, there are no partners; instead,
there is a self-action of correlated fields. The non-Newtonian self-tensions in (7) 
along spatial directions,  $-c^4 \nabla_\nu B^\nu_i/8\pi G \equiv - c^4(\nabla_\nu \nabla_i a^\nu  +  \nabla_\nu \nabla^\nu a_i)/8\pi G$, prevent the centripetal collapse of inhomogeneous energy profiles with
  $(\partial_i B) c^4/16\pi G $ $\neq 0$.  
   The covariant counterbalance for non-equilibrium oscillations around equilibrium profiles and for stationary organisations can be easily verified in (7)-(8) under the static self-organization of relativistic 4-velocities $c u_{\mu} \equiv g_{\mu \nu} c u^{\nu} \equiv$ $\left\{c g_{o} \gamma ;-c \gamma\left(\beta_{i}-g_{i}\right) / g_{o}\right\}\Rightarrow \left\{c \sqrt{g_{o o}} ; 0\right\}$ and GR 4-accelerations $c^2a_{\mu} \equiv c^2g_{\mu \nu} a^{\nu} \equiv c^2u^{\nu} f_{\nu \mu}$ $\Rightarrow\left\{0 ;-c^{2} x_{i} r_{o} /\left[r^{2}\left(r+r_{o}\right)\right]\right\}, a^{\mu}=\left\{0 ;+c^{2} x^{i} r_{o} /\left[r^{2}\left(r+r_{o}\right)\right]\right\}$ of the generated mass-energy densities in (9) with the metric consequences $g_{o o}=1 / g_{o o}=r^{2} /(r+$ $\left.r_{o}\right)^{2}, g_{o i}=g^{o i}=0, g_{i j}=-\delta_{i j}, g^{i j}=-\delta^{i j}$.
  
  The monistic framework presented elucidates how adaptive time rates and nonlocal correlations can explain phenomena traditionally handled within a Newtonian dualism, while simultaneously preserving the integrity of the spatial geometry. The new theoretical construct opens avenues for understanding cosmic dynamics without relying on the conventional partnership of interacting bodies in dualistic frameworks. The conceptual monism of continuous material organizations can potentially guide future research in geometrodynamics for the nonlocal organization of self-gravitating systems.

 \section { Shannon information in the Yin-Yang energy distribution with zero volume integral }

\subsection {Static equilibrium of Euclidean matterspace}

The stabilizing balance of internal counter-stresses in the extended charge was argued by Poincaré [17]. Our physical interpretation of the contracted Bianchi identities (7) or (8) 
through the dialectical unity and struggle of kinetic repulsions and metric attractions corresponds not only to Poincaré's guidance but also agrees with the stable cosmology of Kant. The latter was supported by the Hegelian ideas for the primacy of kinetic repulsion in the visible existence of volumetric matter (without the consequences of gravitational collapse or the Beantley paradox). Repulsion provides matter for Newtonian attraction, as Hegel explained from the dialectical stability of material profiles on Earth and in the cosmos. Below we decipher the Hehelean balance of repulsion and attraction in mathematical terms of positive (yang, active) and negative (yin, reactive) energy densities in the Shannon information continuum with the zero integral (5). 

The Lagrange equation of motion (7) 
 for the temporal axis $(\mu \Rightarrow 0)$ in the flat-space physics (4) with $B_{o}^{i}=0, B_{i}^{o} \neq 0$ and $B_{o}^{o}=B_{i}^{i}=B / 2>0$ can be expressed as follows:
\begin {eqnarray} \label {eq:1041} 
\left\{\begin{array}{l}
	\nabla_{o} \frac{B}{2} \equiv \partial_{o} \frac{B}{2}=\nabla_{o} B_{o}^{o}+\nabla_{i} B_{o}^{i} \equiv\left(\partial_{o} B_{o}^{o}+\Gamma_{o \lambda}^{o} B_{o}^{\lambda}-\Gamma_{o o}^{\lambda} B_{\lambda}^{o}\right)  
	\\
	+\left[\partial_{i} B_{o}^{i}+\Gamma_{i \lambda}^{i} B_{o}^{\lambda}-\Gamma_{i o}^{\lambda} B_{\lambda}^{i}\right]=\left(\partial_{o} \frac{B}{2}-\Gamma_{o o}^{j} B_{j}^{o}\right)+\left[\Gamma_{i o}^{i} B_{o}^{o}-\Gamma_{i o}^{j} B_{j}^{i}\right], \\
	\Gamma_{i o}^{i} B_{j}^{j}=\Gamma_{i o}^{i} B_{o}^{o}=\Gamma_{i o}^{j} B_{j}^{i}+\Gamma_{o o}^{j} B_{j}^{o} \equiv \Gamma_{\nu o}^{j} B_{j}^{\nu}=\Gamma_{\nu o}^{\mu} B_{\mu}^{\nu}
\end{array}\right.
\end {eqnarray}
These dynamical relations can be read as the bounds for admissible 3-potentials $g_{i} \equiv g_{o i} / g_{o}$ in the time-varying condition $\Gamma_{i o}^{i}=2 \Gamma_{\nu o}^{\mu} B_{\mu}^{\nu} / B$ of the flat matterspace. Recall from (6) that $\Gamma_{i o}^{j}=\Gamma_{i o}^{i}=0$ in the absence of gyro-potentials $g_{1}=g_{2}=g_{3}=0$, wich leads to $B_{j}^{o}=0$ for such laminar flows or static states without rotations.

Geometrodynamic equations (8) 
for three spatial axes, when $\mu \Rightarrow i$ in (7), 
can be simplified for the static equilibrium where one can use $g_{i}=0, g^{o i}=$ $0, \Gamma_{i o}^{j}=\Gamma_{i k}^{j}=0, a_{o} \equiv u^{i} f_{i o}=0, a^{o} \equiv g^{o \nu} a_{\nu}=0, \nabla_{i} a^{j}=\partial_{i} a^{j}, a_{i}=-\partial_{i} \ln g_{o}(\boldsymbol{x})=$ $\partial_{i} \ln \left[1 / g_{o}(\boldsymbol{x})\right], a^{j}=g^{j i} a_{i}=\delta^{j i} \partial_{i} \ln g_{o}(\boldsymbol{x}) \equiv \partial^{j} \ln g_{o}(\boldsymbol{x}):$
\begin {eqnarray} \label {eq:1142}
\left\{\begin{array}{l}
	\nabla_{\nu} T_{i}^{\nu}=\frac{1}{g_{o}} \partial_{j}\left(g_{o} T_{i}^{j}\right)\\ =\frac{\varphi_{G}^{2}}{8 \pi c^{2} g_{o}}\left\{\partial_{i}\left[g_{o}\left(2 \partial_{j} a^{j}\right)\right]-\partial_{j}\left[g_{o}\left(\partial_{i} a^{j}-\partial^{j} a_{i}\right)\right]\right\}=0, \\ 
	\frac{\varphi_{G}^{2}}{4 \pi}\left[\left(\partial_{i} \ln g_{o}\right) \partial_{j} \partial^{j} \ln g_{o}+\partial_{i} \partial_{j} \partial^{j} \ln g_{o}\right] \\ =\frac{\varphi_{G}^{2}}{4 \pi}\left[\left(\partial_{j} \ln g_{o}\right) \partial^{j} \partial_{i} \ln g_{o}+\partial_{j} \partial^{j} \partial_{i} \ln g_{o}\right], \\
	\frac{\varphi_{G}^{2}}{4 \pi}\left(\partial_{i} \ln \frac{1}{g_{o}(\boldsymbol{x})}\right) \partial_{j} \partial^{j} \ln \frac{1}{g_{o}(\boldsymbol{x})}=\frac{\varphi_{G}^{2}}{4 \pi}\left(\partial_{j} \ln \frac{1}{g_{o}(\boldsymbol{x})}\right) \partial^{j} \partial_{i} \ln \frac{1}{g_{o}(\boldsymbol{x})}, \\
	{\rm or }\ a_{i} \partial^{j} a_{j}=a_{j} \partial^{j} a_{i}, \partial^{j} \equiv \delta^{j k} \partial / \partial x^{k}.
\end{array}\right.
\end {eqnarray}
Here we use the mathematical identity $\partial_{i} \partial_{j} \partial^{j} \ln g_{o}(\boldsymbol{x}) \equiv \partial_{j} \partial^{j} \partial_{i} \ln g_{o}(\boldsymbol{x})$ for the partial derivatives of the logarithmic function $\ln g_{o}(\boldsymbol{x})$ in Euclidean 3-space $\boldsymbol{x}\Rightarrow x^i$. The static matterspace with the spherical symmetry has the rectungular coordinate acceleration $c^{2} a_{i}=-G m x^{i} /\left[r^{2}\left(r+r_{o}\right)\right]$~\cite{Bul2008}, which corresponds to the non-collapsing equilibrium (11). 
It turned out that this spatial acceleration quantitatively coincides with the Newtonian field references at large distances from the origin, when $|\mathbf{x}|=r \gg r_{o} \equiv G m / c^{2}$. But how to find strong field solutions of (11) 
 or self-accelerations $a_{i}$ in corelated matterspace with multiple vertices of mass densities?

\subsection {Shannon information references nullify energy integral in equilibrium self-assembly}

Now we compare the spherical field potential $W(r)=\varphi_{_G} \ln g_{o}(r)=-\varphi_{_G} \ln [1+\varphi_{_G}^{-1}\left.(\sqrt{G} m / r) \right]$ for the interaction strength $-\partial_{r} W(r)$ exerted on probe (passive) charges $\sqrt{G} m_{p r}$ with the Shannon~\cite{Sha1949} transmission capacity $C=w \log [1+$ $\left.\left(w_{o} / w\right)\right] \approx-w \log \left(w / w_{o}\right)$, where the noise power is proportional to the information bandwidth $w$. The universal potential $\varphi_{_G} \equiv c^{2} / \sqrt{G}$ acts as a transmission band for energy potentials $\sqrt{G} m_{k} / r_{k}$ from elementary mechanical charges $\sqrt{G} m_{k}$. One can construct the metric component $g_{o o} \equiv g_{o}^{2}$ and the interaction potential $W(\boldsymbol{x})=\varphi_{_G} \ln g_{o}(\boldsymbol{x})$ in any point $\mathbf{x}$ of the continuous matterspace with many information and energy centers in a line of the Shannon theory:
\begin {eqnarray} \label {eq:1243}
\begin{array}{r}
	\frac{1}{g_{o}(\boldsymbol{x})}=1+\frac{r_{1}}{\left|\boldsymbol{x}-\boldsymbol{\xi}_{1}\right|}+\frac{r_{2}}{\left|\boldsymbol{x}-\boldsymbol{\xi}_{2}\right|}+\ldots+\frac{r_{n}}{\left|\boldsymbol{x}-\boldsymbol{\xi}_{n}\right|}, r_{k} \equiv \sqrt{G} m_{k} / \varphi_{_G}, \\
	W(\boldsymbol{x})=-\varphi_{_G} \ln \left(1+\frac{1}{\varphi_{_G}} \sum_k \frac{\sqrt{G} m_{k}}{\left|\boldsymbol{x}-\boldsymbol{\xi}_{k}\right|}\right), \\
	\varphi_{_G} \boldsymbol{a}(\boldsymbol{x}) \equiv-\boldsymbol{\partial} W(\boldsymbol{x})=-\varphi_{_G} \frac{\frac{r_{1}\left(\boldsymbol{x}-\boldsymbol{\xi}_{1}\right)}{\left|\boldsymbol{x}-\boldsymbol{\xi}_{1}\right|^{3}}+\frac{r_{2}\left(\boldsymbol{x}-\boldsymbol{\xi}_{2}\right)}{\left|\boldsymbol{x}-\boldsymbol{\xi}_{2}\right|^{3}}+\ldots+\frac{r_{n}\left(\boldsymbol{x}-\boldsymbol{\xi}_{n}\right)}{\left|\boldsymbol{x}-\boldsymbol{\xi}_{n}\right|^{3}}}{1+\frac{r_{1}}{\left|\boldsymbol{x}-\boldsymbol{\xi}_{1}\right|}+\frac{r_{2}}{\left|\boldsymbol{x}-\boldsymbol{\xi}_{2}\right|}+\ldots+\frac{r_{n}}{\left|\boldsymbol{x}-\boldsymbol{\xi}_{n}\right|}}. 
\end{array}
\end {eqnarray}

The instantaneous transfer of big data is embedded in a multi-vertex matterspace with the correlated information and mass-energy distribution. The
non-locality of macro-organisations is difficult to explain by the empty-space model of dualistic approaches, which deny the continuous matterspace of Cartesian physics.

Now one can relate active and passive mass densities with the Shannon potential $W(\boldsymbol{x})$ for energy interactions and information transfer:
\begin {eqnarray}
	\cases {
		{\sqrt G} \mu_p({\bm x})  \equiv \frac {[-{\bm \partial }W({\bf x})]^2}{4\pi\varphi_{\!_{_G}} } = 
		\frac {  {\bm \partial }^2 W({\bf x})} {4\pi} \equiv  {{\sqrt G}\mu_a({\bf x})},
		\cr 
		\int\!d^3x\! {\sqrt G}\mu_a({\bf x})  \!\equiv
		\!\int\! \frac { \!d^3x \!} {4\pi}{\bm \partial }^2 W({\bf x})\!=\! \!\int\!
		\frac { \!d{\bm S} \!} {4\pi}
		{  {\bm \partial } W({\bf x})} \! =\!{\varphi}_{\!_{_G}}\! \sum_{k}\! r_k \!\equiv \! {\sqrt G} \sum_{k}\! m_{k},
		\cr
		\int\!\! {\sqrt G}\mu_p({\bf x}) d^3x \!\equiv \!  \frac {{\varphi}_{\!_{_G}} }{4\pi }\!\! \int\!\! 
		\left (\!
		\frac {    \frac {{({\bf x} - {\bm \xi}_1)r_1}}{|{\bm x} - {\bm \xi}_1 |^3}    + \frac {({\bf x} - {\bm \xi}_2)r_2}{|{\bf x} - {\bm \xi}_2 |^3} +...+ \frac {({\bf x} - {\bm \xi}_n)r_n}{|{\bf x} - {\bm \xi}_n |^3}}{ 1 + \frac {r_1}{|{\bm x} - {\bm \xi}_1 |}     + \frac {r_2}{|{\bf x} - {\bm \xi}_2 |} +...+ \frac {r_n}{|{\bm x} - {\bm \xi}_n |} } \!\right )^{\!2}\!\!\!d^3x 
		= \! {\sqrt G}\sum_{k}\! m_{k},
		\cr
		\int\!\! {\sqrt G}\mu_p(\!{\bf x}\!) W\!(\!{\bm x}\!) d^3x \!\equiv \!
		-\!\! \int\!\! \frac { d^3x }{4\pi}\!\!
		\left (\!\!
		\frac { \varphi_{\!_{_G}}\!\!  \sum_k\!\! \frac {   {({\bf x} - {\bm \xi}_k)r_k}}{|{\bm x} - {\bm \xi}_k |^3}  
		}{ 1 + \sum_k\!\!\frac {r_k}{|{\bf x} - {\bm \xi}_k |}  } \!\!\right )^2\!\!\!\!ln\! \left( \!\!1\! + \!    \sum_k
		\!\!  \frac {r_k}{|{\bf x} - {\bm \xi}_k |} 
		\!\! \right )\! =\!  -\!c^2 \!\sum_{k}\! \!m_{k},
		\cr \int\!d^3x [{\sqrt G}\mu_a({\bm x})\varphi_{\!_{_G}}  
		+   {\sqrt G}\mu_p({\bm x}) W\!({\bm x})] = 0. 
}
\end {eqnarray}
Here the positive rest-energy of active emergent charges $\varphi_{G}\!\sum_k\! \sqrt{G} m_{k} \equiv \sum_k m_{k} c^{2}$ > 0 is compensated by the volumetric integral of the negative potential energies $\sqrt{G} \mu_{p}(\boldsymbol{x}) W(\boldsymbol{x})<0$ of the passive densities $\sqrt{G} \mu_{p}(\mathbf{x})$ in the Shannon interaction potential from (12). This yin-yang dialectic with the zero energy balance of non-local self-assemblies underlines how informatics can be an integral part of understanding the correlated interplay of distributed energies, forces and big data transmissions within an isolated material system.

The appearance of active charge densities $\sqrt{G} \mu_{a}(x)$ requires metric stresses and local energy costs for the generation of rest energy densities $\mu_{a}(x) c^{2}>0$ due to the positive self-potential $\varphi_{G} \equiv c^{2} / \sqrt{G}$. Positive rest energy of macroscopic bodies can be considered as kinetic and related to microscopic motions or fast micro-oscillations. The passive (or reactive) charge density $\sqrt{G} \mu_{p}(x)$ is affected by the Shannon potential $W(x)<0$ for information rates and metric correlations. The local energy imbalance $\mu_{a}(x) c^{2}+\sqrt{G} \mu_{p}(x) W(x) \neq 0$, where $\mu_{a}(x)=\mu_{p}(x)$, occurs under the volumetric zeroing in  (13) 
 of active (yang, kinetic) and reactive (yin, potential) self-energy integrals in the non-local distribution of matterspace with metric stresses and information exchange.

The Yin-Yang worldview has a long history of studies in the East. In Europe, the most consistent proponent of the complete cancellation of the rest energy of a self-gravitating star by its internal binding energies was  Pascual Jordan. From the observed coexistence of old and new stars he deduced their spontaneous formation with zero summary energy~\cite{Jor1939}. At that time, the gravitational binding energy in the strong-field regime was not modelled in general relativity. And the advanced hypothesis  of \lq a new and constant production of mass in space\rq{} with zero relativistic energy was quite puzzling.

Monistic geometrodynamics, where the flat matterspace of a quasi-isolated system relates the self-generated mass densities to local accelerations in the correlated distribution, returns to the integral cancellation of kinetic and potential self-energies under equal active and passive masses. The equivalence of passive (reactive) and active (attractive) mass densities is a mathematical consequence
of four variational equations $T_{o}^{\mu}(x)=0$ for the flat-space continuum of massive fields under the adaptive local time. The vector geometrodynamics of the correlated tensor densities $T_{i}^{\mu}(x) \neq 0$ in the nonlocal mass-energy organization obey the Bianci identities for relativistic four-accelerations. The latter are the primary reason for the observable existence of the physical world with inertial masses and adaptive changes of their densities.

The field monism of living matter, material thoughts and inert or toxic masses in the common energetic and informational filling (13) 
 of Vernadsky\rq{s} Geo-Bio-Noosphere does not allow, in principle, to give an independent definition of neither living nor inert (inanimate) matter in a satisfactory way. Due to the gloval superposition in common energy and information content, we can think about the simultaneous emergence of living and inert hierarchies after the Big Bang or the eternal existence of living self-organisations in bouncing models of the Metagalaxy.

\section{Experiment can falsify theory, never validate it}

\subsection{Misinterpretation of world nonlocality in local measurements of energy-momentum exchange}

According to Gödel\rq{s} theorems, it is impossible to establish a closed algebraic theory without reference to external mathematical constructions. Similarly, there are no completely isolated subsystems in their hierarchical immersion in the physical whole. Nevertheless, for many computational models it is more convenient to consider a quasi-isolated subsystem as an isolated energy body or an elementary particle. For relatively weak mechanical relations between formally isolated masses, such an approximation of reality suggests the field concept of mutual forces between inertial partners. Such historical division of the monistic whole into almost isolated dense regions of the common space of nonlocal mass, energy and information led first to the modeling success of the two-body problem, and then to the triumph of statistical ensembles in dual partitions of the continuous medium into atoms and molecules.

The dualistic worldview not only allowed for the discovery of Neptune at the tip of a pen, but also led to the theoretical standards in simplified physics of massive particles and massless fields modeling long-range interactions. It is believed that since the Standard Model has allowed experiments with highenergy particles to be described with the highest modern precision, monistic alternatives are unable to compete with references to Newtonian mechanics and gravitation on the mega- and macroscale. It is often claimed by Cartesian opponents that since the main predictions of special and general relativity have been quantitatively confirmed in practice, these coherent extensions of Newtonian mechanics and gravitation also preserve the dual concept of localized partners in the ambient
 interaction field. 

On the contrary, Kuhn claimed that \lq\lq Einstein theory could be accepted only with the recognition that Newton\rq{s} was wrong\rq\rq{}~\cite{Kuh1962}. Indeed, Einstein\rq{s} geodesic motion~\cite{Ein1914} has four mechanical potentials $g_{\mu}$, while incomplete classical mechanics accepts only one gravitational potential $g_{o}$; masses after warming weigh more only in Einstein\rq{s} physics, etc. After decades of explicit theoretical concerns about the accepted interpretation of General Relativity~\cite{Bri1978,Nar1985}, cosmologists still believe in \lq\lq the well known fact that the weak field, low velocity, low pressure corresponding limit of GR is Newtonian gravity, as evidenced by solar system tests\rq\rq{}~\cite{Row2015}. 

The true fact is that the solar system exists in few billions of years, while the dualistic physics of local Newtian masses can explain its mechanical stability only for a few handred years~\cite{Lap1975}. Instantaneous geometric correlations for self-consistent stresses and mass densities of a nonlocal medium or elastic matterspace in geometrodynamics (7)-(8)
can preserve the dynamical self-assembly of the solar system indefinitely if the dissipative exchange with nearby stars is neglected. The spherically symmetric continuum of mass in the non-locally organized matterspace of a secluded star admits not only instantaneous correlations of metric stresses, but also diverging-converging spherical waves of mass density of longitudinal character. Such a cosmic coherence of extended matter around the dense visible densities of the shining star explains three signals in Kozyrev\rq{s} telescopic experiments, including the wave echo from the periphery to the center, often interpreted as an \lq advanced\rq{} flux to the future positions of a moving star~\cite{Koz1976,Lav1992}. 

After the numerical study of the appearing  \lq advanced\rq{} - retarded fronts of converging - diverging longitudinal shifts in the monistic space continuum, one can, for example, predict future locations of earthquakes or weather disasters. Such promising applications of the geometrodynamic equation (7) to the future energy events in the nonlocal Earth continuum or to the medium-term prognosis for the Sun\rq{s} activity~\cite{Kor2020} can be a practical reason for further studies of monistic macrophysics with instantaneous entanglement of stresses and \lq advanced\rq{} returns of bound longitude perturbations or wave ekhos. Geometrodynamics of elastic matterspace can be universally applied to as to megascale self-oscillations in cosmological formations and pulsating stars, as to force-free \lq zitterbewegung\rq{} of microscale of microscale densities~\cite{Sch1938,Zhi2008}.  

Non-local auto-organization of correlated densities of an isolated system at both scale limits is consistent with self-interaction under conservation of mechanical energy. Such autodynamics does not obey Newtonrq{s} 2nd law for change of energy-momentum of a considered body by an external force without feedbacks. Local measurements detect apparent energy changes for probe elements, and such dissipative changes are used to misinterpret the conceptual monism of elastic matterspace in dualistic terms of hierarchical partners with individual energy contents.

\subsection{Schwartzschild metric with model singularities has not yet been falsified by measurements}
Einstein performed a thought experiment in 1939 and logically rejected the Schwarzschild metric solution in the strong field region together with unphysical singularities~\cite{Ein1939}. The introduction of Mach\rq{s} nonlocal ideas into Einstein\rq{s} tensor formalism allowed in 2007 to rewrite the 4 -interval into the non-Schwarzschild version~\cite{Bul2008}, $d s^{2}=\left[r d x^{o} /\left(r+r_{o}\right)\right]^{2}-d r^{2}-r^{2}\left(d \theta^{2}+\sin ^{2} \theta d \varphi^{2}\right)$, which quantitatively justified the equivalence of active and passive masses in Euclidean matterspace.

It seems that the December 1915 metric~\cite{Sch1916} is still forbidden for criticism because it is identified by the mass media as the most famous achievement of Einstein, not   Schwarzschild. Theorists of leading gravitational journals try to protect the uniqueness of the Schwarzschild metric by Birkhoff\rq{s} theorem. At the same time they ignore the fact that this theorem is valid only for the void and not for nonlocal reality with Cartesian matter extension. Experimentalists are also convinced that \lq the reluctant father of black holes\rq{}  was wrong to deny the existence of metric singularities in 1939, since the predictions of the Schwarzschild metric have been very reliably verified by numerous measurements. However, any mathematician can easily check that the 2007 metric for the nonlocal filling of Euclidean space by radial masses with equal passive and active densities leads to the same~\cite{Bul2012} quantitative predictions as the Schwarzschild modelling of solar fields in well-known GR probes. Remember that measurements do not validate competing theories, but only falsify incorrect approaches as accuracy increase~\cite{Pop1959}. The strong field predictions from both the 1915 and 2007 4-intervals have not yet been falsified in the available range of measurements.

The second order gravitational redshift can in principle distinguish between curved empty space in dualistic physics and flat matterspace with adaptive time in our monistic alternative. Again, the 3D Euclidean sub-interval for Cartesian matter-extension with $g_{o o}^{2007}=r^{2} /\left(r+r_{o}\right)^{2}$ for the static radial organization maintain the pseudo-Riemannian 4D geometry and the same first-order corrections to Newtonian gravity for Mercury\rq{s}  perihelion precession, redshift, and light bending as predicted by the Schwarzschild metric with $g_{o o}^{1915}=1-2 r_{o} c^{2} / r$ for the main GR tests in the weak negative potential $\tilde{\varphi}=-r_{o} / r$, when $r_{o} \equiv$ $G m / c^{2} \ll r$. But the second order relativistic corrections are different for the warped empty space and for the Euclidean material space. Precise measurements of the physical time dilation in the Earth-Sun-Moon system with the time-varying weak potential $\tilde{\varphi}$ make it possible to distinguish the 1915 time dilation $(d \tau-d t) / d t \equiv \sqrt{g_{o o}^{1915}}-1 \approx \tilde{\varphi}-\left(\tilde{\varphi}^{2} / 2\right)$ in curved empty space from alternative time dilation $\sqrt{g_{o o}^{2007}}-1 \approx \tilde{\varphi}+\tilde{\varphi}^{2}$ in the flat matterspace.

\subsection{3D space flatness from SQUID measurements}
So far inhomogeneous gravity and acceleration have never altered SQUID records for macroscopic loops of superconductors. Such null results are consistent with the macroscopic flatness of the laboratory space with the highest quantum accuracy~\cite{Bul2009}. If anyone were to measure the contribution of gravitational fields or accelerations to the quantization of magnetic flux from self-correlated currents in superconducting contours, the Euclidean geometry of the laboratory space would in principle be falsified. In turn, the experimental independence of the quantized magnetic flux from changes in the gravitational field or changes in the acceleration changes falsifies its theoretical relation to the 3 -space curvature in dualistic interpretations of Einstein\rq{s}  geometric fields. SQUID-based accelerometers could only exist in a metrically warped 3 -space and have never been built in physical reality. Based on the absence of SQUID accelerometers and gravimeters, new generation of relativists should think about the metricdependent 4-interval with a universal (Euclidean) 3-section and adaptive local time in Einstein\rq{s}  physics.

\subsection{Electrodynamic experiments can falsify the dogma of retarded actions \lq from there to here\rq{}}
The GR metric forces are definitely not the pulling \lq action-at-a-distance\rq{.} They seem to be local geometric pushes, like superpenetrating liquid pushes in Lomono\-sov\rq{s}  aetheric model. The pan-unity of visible and invisible matter in the Lomono\-sov and Umov continuum of kinetic energy with inertial heat means, in modern terms, spatial non-locality of inertial densities with correlated field stresses from General Relativity. By interpreting the mathematical beauty of geometric fields Einstein has publicly admitted the ponderable ether: \lq\lq ... space is endowed with physical qualities; in this sense, therefore, there exists an ether. According to the general theory of relativity space without ether is unthinkable; for in such space there not only would be no propagation of light, but also no possibility of existence for standards of space and time (measuring-rods and clocks), nor therefore any space-time intervals in the physical sense. But this ether may not be thought of as endowed with the quality characteristic of ponderable media, as consisting of parts which may be tracked through time. The idea of motion may not be applied to it\rq\rq{}~\cite {Ein1920}.

Time-varying stresses are simultaneous in the nonlocal field theory, maintaining constant energy integrals at all world moments $x^{o}$. Wave modulations of correlated densities can be retarded (diverging from the mass vertex) or \lq advanced\rq{} (converging to it). The monistic GR formalism for continuous matterspace with dense (visible) and rarefied (invisible) regions may shed some light on the large-scale correlation of galaxy motions~\cite{Mys}, the megaparsec alignment of the pulsars~\cite{Ali}, and the macroscopic entanglement of photons in the Earth experiments commented on for the 2022 Nobel Prize in Physics.

The instantaneous action on the telescope detector~\cite{Koz1976,Lav1992} under the actual direction to a moving star suggests timeless Newton and Coulomb forces without temporal retardation due to the intuitive \lq speed of gravity\rq{,} estimated by Laplacce above $6 \times 10^{6} c$~\cite{Lap1975}. The Coulomb\rq{s} forces have indeed exhibited the rigid motion with high-energy beams of ultrarelativistic electrons~\cite{San2015}. These well-executed experiments at the Frascati National Laboratory are consistent in principle with the monistic concept of nonlocal matterspace. The modified scheme with high-enegy beams of relativistic charges was also proposed~\cite{Bli2018} to falsify the empty space model beyong reasonable doubts. A clear rejection of conceptual exprements can provide for the competing worldviews of monists and dualists. The non-local reading of continuous electric charges and mechanical masses supports the local self-action mechanism for the inverse square forces~\cite{Bul2023}. In such monistic physics, correlated geodesic accelerations push inertial densities \lq from here to there\rq{} instead of the classical pulling \lq from there\rq{,} as Newtonian action-at-a-distance states.

Since local self-accelerations define the long-range interactions in the nonlocal self-assembly of matterspace, the intervention of external forces in local regions of such self-assembly resonating with auto-oscillations can perturb the correlated dynamics of distant densities. Non-metric external forces perturb the time-averaged equilibrium of the nonlocal system with the metric energy-information potential $\mathrm{W}$, which in (13) 
leads to the mutual compensation of the kinetic (yang) and metric (yin) energy integrals. Such a spooky action at a distance by a non-locally organised system was originally predicted from the paradoxes of quantum mechanics with instantaneous reduction of the wave function~\cite{Ein1935}. Encountered numbers of non-local laboratory experiments and spooky communications between bio-cells, bacteria, animals and humans are outside the immediate comments of monistic field physics for macro-correlated hierarchies of Shannon information in geometrized mass-energy.

\section {Conclusion} 

Observations of the correlation of cosmic matter on megaparsecs in the tenth decade and the Nobel Committee Prize for macroscopic experiments violating Bell\rq{s} inequality in 2022 have ended conceptual disputes about the exit of quantum nonlocality from the notorious  microcosm to all spatial scales. Renewed interest in the monistic Cartesian matter-extension and field theories of continuous mass distribution has prompted a covariant generalisation of previously geometrized  matter for static equilibrium \cite{Bul2008}. 
	
	The formal division of a highly inhomogeneous medium into localised carriers and virtually empty space has led to very successful mechanical and electrical models with remote interaction between the designated partners. 
 The calculation of model thrusts \lq from there to here\rq{} for massive points using the geometric law of inverse squares is much easier than the estimation of non-local self-accelerations \lq from here to there\rq{} in a non-local medium with timeless stresses \cite{Bul2023}. The simplified modelling always has its price - the empty space without elasticity cannot lead to the time-averaged equilibrium of self-gravitating systems and gives rise to many puzzles in astronomical observations, starting with the Bentley\rq{s} paradox.

 Local dissipation processes from external (non-metric) interventions or intensive radiation losses of  dense elastic regions violate the metric organization with nonlocal correlations in closed mechanical systems. Non-metric energy-momentum exchanges with $\nabla_\mu j^\mu (x)\neq 0$ are much easier to analyze in the dissipative terms of the  Newton Second Law, rather than in the Cartesian terms of integral energy-momentum exchanges  with continuous vortex densities. The dualistic reinterpretation of purely field matter has had tremendous quantitative success for practice, leading to the Standard Model of physics for separated particles and their field mediators. Despite the philosophical directives and the predicted evolution of physics towards the nondual field theory \cite {EI1938}, one can say again that \lq\lq  people
slowly accustomed themselves to the idea that the physical states of space itself were the finite physical reality\rq\rq{.} Cartesian publications on matter-extension and monistic physics are  still underrepresented in the current volume of dualistic alternatives for mechanical events. 

The correlated matter-space or elastic space-plenum of the ancient Greeks can have many unique applications. It consistently accepts the nonlocal nature of elastic self-organization both in the quantum microcosm and in the whole Metagalaxy.  Material space-plenum can be continuously filled with bits of information, which cannot be done for emptiness. Therefore, monistic physics has the power to initiate a new informatics of big data based on instantaneous stresses of the elastic cosmos, but not only on the slow wave traffic of dissipative exchanges with retardation. 

 Practical applications of the monistic picture and the geometrodynamics of correlated distributions in macroscopic physics, biology, informatics, cosmology,  hydraulic engineering and many other energy-related disciplines go  beyond the presented extension of the static self-assembly to  its covariant  self-motion in the absence of non-metric  forces. 
 Finally, the twentieth-century initiatives of Einstein-Grossmann and Mie should one day reinforce  the flat matter-space-plenum of the ancient Greeks, the vortex matter-extension of Descartes, and the etheric  world pan-unity in the monistic terms of Russian Cosmism. The renaissance of Euclidean 3-space for  correlated inertial fields addresses the discovered entanglement of macroscopic states and bridges relativistic physics with quantisation, advancing geometrodynamics to decipher the non-local transfer of big data in the material field reality.

 \bigskip \noindent 
 {\textbf {Acknowledgments}}. 
 The author is grateful to Prof. Guido Pizzella for visiting Frascati National Laboratories to study the details of the conceptual experiment [34].






\begin{thebibliography}{}
		
		\bibitem{Ari} J. Fritsche, \emph{Aristotle on Space, Form, and Matter} (Aristotle, Physics IV:2, {\bf 209} B 17), \emph{Archiv für Begriffsgeschichte}
	{\bf 48} (2006) 47.
	
	
	
	
	
	\bibitem{Des}
	D. Garber,  \emph{Descartes\rq{} Metaphysical Physics}, University Chicago Press (1992).
	
	\bibitem{Des1} \emph{ Descartes\rq{} Physics}, in {The Cambridge Companion to Descartes}, ed. Cottingham, J.,  
	Cambridge University Press, New York (1992). 
	
	
	\bibitem{Lom} M.V. Lomonosov,  \emph{Notes on the severity of bodies}, 	Complete Works, 11 Vols., 1743, Vol.2,
	eds. S. Vavilov and T. Kravetz, Akad. Nauk. SSSR, Moscow and Leningrad (1950).
	
	
	\bibitem {Umo} N.A. Umov,  \emph{Beweg-Gleich. d. Energie in contin. Korpern. Schomilch.}, Zeitschriff d. Math. und Phys.  {\bf XIX} (1874). 
	
	
		
		
		\bibitem{Mie}	 G. Mie, \emph{Grundlagen einer theorie der materie.} 
		\emph{Ann. der Physik} {\bf 37} (1912) 511 and {\bf 39} (1912) 1. 
		
		
		\bibitem{Ein1920} A. Einstein,  \emph{Ether and the Theory of Relativity}, Lecture at the University of Leiden on 5 May 1920,  published in English  by Methuen \& Co. Ltd, London (1922).
		
		
		\bibitem{Ali} D. Hutsem{\'e}kers, L.  Braibant, V.   Pelgrims, D. 
		Sluse, \emph{   Alignment of 			quasar polarizations with large-scale structures,}
		\emph{Astron. \& Astrophys.} {\bf A18} {(2014)} 572.
		
		
		\bibitem{Mys}  J.H. Lee, M.  Pak, H.   Song, H.-R.  Lee, S.  Kim, H. Jeong,
		\emph{Mysterious 			coherence in several-megaparsec scales between galaxy rotation			and neighbor motion},
		\emph{Astrophys. J.} {\bf 884}(2) (2019) 104.
		
		
		\bibitem{Ein1914} 	 A. Einstein, \emph{Die formale Grundlage der allgemeinen Relativitatstheorie}, 
		\emph{Sitzungsber. preuss. Acad. Wiss.} {\bf 2} (1914) 1030. 
		
		
		
		
		
		
		
		
		
	
	\bibitem{Hil1915} G. Hilbert, \emph{Die Grundlagen der Physik  (Erste Mitteilung)}, \emph{Nachrichten von der Gesellschaft der Wissenschaften zu Göttingen, Mathematisch-Physikalische Klasse }  (1915) 395.  
	
	
	\bibitem{Ein} A.	Einstein, \emph{Die Grundlage der allgemeinen Relativitatstheorie},   
	\emph{Ann. Phys.} {\bf 49} (1916) 769
	
	
	
	
	\bibitem{Pop1959} K. Popper, \emph{Logik
		der Forschung}, Vienna  (1934), trans.  
	\emph{The Logic of Scientific Discovery} (1959).
	
	
	\bibitem{Ein1939}
	A. Einstein, \emph{On a Stationary System With Spherical Symmetry
		Consisting of Many Gravitating Masses}, \emph{Ann. Math.}
	\emph{40} (1939) 922.
	
	
	\bibitem{EI1938} A. Einstein, L. Infeld, \emph{The Evolution of Physics}, ed. by C.P. Snow, Cambridge University Press (1938).
		
		
		
	
	
	
	\bibitem{Bul2008}  I.{\'E}. Bulyzhenkov-Widicker,  \emph{Einstein\rq{s} Gravitation for Machian Relativism of Nonlocal Energy-Charges},
	\emph{Int. J. Theor. Phys.} {\bf 47} (2009) 1261.
	
	\bibitem{Poi} H. Poincar{\'e}, \emph{Sur la dynamique
		de l\rq{\'e}lectron} (received on 23 July,1905). Rendiconti del Circolo Matematico di Palermo {\bf 21} (1906)
	129. In English: 
	\emph{On the dynamics of the electron}.
	{Rendiconti del Circolo Matematico di Palermo} {\bf 21} (1906) 129.
	
	
	
	\bibitem{GE1912}  A. Einstein, M.  Grossmann, \emph{
		Entwurf einer verallgemeinerten Relativitätstheorie und einer Theorie der Gravitation},\emph{Zeitschrift für Mathematik und Physik} {\bf 62} (1913) 225. 
	
	
	
	
	
	\bibitem{Sha1949} C.E.	Shannon,  
	\emph{Communication in the presence of noise},  
	\emph{Proc. Institute of Radio Engineers}  {\bf 37}(1) (1949) 10.
	
	
	\bibitem{Jor1939} P. Jordan, \emph{Bemerkungen zur Kosmologie}, \emph{Ann. Phys.} {\bf 428} (1939)
	64. 
	
	\bibitem{Kuh1962} T.S. Kuhn, \emph{The Structure of Scientific Revolutions} (Chicago,
	1962).
		
	
	\bibitem{Bri1978}  L. Brillouin, \emph{Relativity reexamined}, New York and London,
	Academic Press (1970).
	
	\bibitem{Nar1985} J.V. Narlikar, \emph{A Random Walk in General Relativity and
		Cosmology}, ed. by N.K. Dadhich, J. Krishna Rao and C.V. Vishveshwara
	(Wiley Eastern, New Delhy, 1985).
	
	\bibitem{Row2015}  R.D. Rowland, \emph{On claims that general relativity differs from
		Newtonian physics for self-gravitating dusts in the low velocity, weak
		field limit},  \emph{Interna-tional Journal of Modern Physics D}   {\bf 24} (08) (2015)
	1550065.
	
	\bibitem{Lap1975}  P.S. Laplace, \emph{The System of the World}, Sagwan Press,
	London (2018); trans. from \emph{Le Systeme du Monde}, Paris (1795).
	
	\bibitem{Koz1976}  N. Kozyrev, \emph{Flare Stars}. Proceedings of the Byurakan Symposium,
	October 5-8, 1976 (Yerevan, 1977, in Russian), p. 209.
	
	\bibitem{Lav1992} M.M. Lavrent\rq{}ev, I.A. Eganova, V.G. Medvedev et al., \emph{On the scanning
		of the stellar sky by the Kozyrev probe}, \emph{Dokl. AN} 323(4) (1992).
	
	
	
		
	\bibitem{Kor2020}  S. Korotaev, N. Budnev, V. Serdyuk, E. Kiktenko, D. Orekhova,
	J. Gorokhov, \emph {Macroscopic nonlocal correlations in reverse time by data
	of the Baikal Experiment}, \emph {J. Phys.: Conf. Ser.} \textbf{1557} (2020) 012026.
	
	
	\bibitem{Sch1938}  E. Schrödinger, \emph { Über die kräftefreie Bewegung in der
	relativistischen Quantenmechanik [On the free movement in relativistic
	quantum mechan- ics]} (in German, 1930), \emph {Sitzungsberichte der
	Preuischen Akademie der Wissenschaften}, \emph {Physikalisch-mathematische
	Klasse} 1 (1930) 418.
	
	\bibitem{Zhi2008}  W. Zhi-Yong and X. Cai-Dong, \emph {Zitterbewegung in quantum field
	theory}, \emph {Chinese Physics} B {\bf 17}(11) (2008) 4170.
	
	\bibitem{Sch1916}  K. Schwarzschild, \emph {On the gravitational field of a mass point
	according to Einstein\rq{s} theory}, \emph {Sitzungsber. Preuss. Akad. Wiss. Berlin
	(Math. Phys.)} (1916) 189.
	
	\bibitem{Bul2012}   I.{\'E}. Bulyzhenkov, \emph {Geometrization of Radial Particles in
	Non-Empty Space Complies with Tests of General Relativity},  \emph {Journal of
	Modern Physics} 3 (2012) 1465.
	
	\bibitem{Bul2009}  I.{\'E}. Bulyzhenkov, \emph {Relativistic quantization of Cooper pairs and
	nonlocal electrons in rotating superconductors}, \emph {Jour. Supercond. Nov.
	Magn.} {\bf 22} (2009) 627.
	
	
		
	
	
	
		\bibitem{San2015}  R. de Sangro, G. Finocchiaro, P. Patteri, M. Piccolo, G.
		Pizzella, \emph { Measuring propagation speed of Coulomb fields},  \emph {The European
		Physical Journal C} \textbf{75} (2015) 137.
		
		\bibitem{Bli2018}  S.V. Blinov,  I.{\'E}. Bulyzhenkov, \emph {Verification of the Rigidity of
		the Coulomb Field in Motion}, \emph {Russian Phys. Jour.} \textbf{61} (2018) 321.
		
		\bibitem{Bul2023}   I.{\'E}. Bulyzhenkov, \emph {Coulomb Force from Non-Local Self-Assembly
		of Multi- Peak Densities in a Charged Space Continuum}, \emph {Particles} {\bf 6}
		(2023) 136.
		
		\bibitem{Ein1935}  A. Einstein, B. Podolsky, and N. Rosen, \emph {Can quantum-mechanical
		descrip-tion of physical reality be considered complete?},  \emph {Phys. Rev.} {\bf 47} (1935)
		777.
		
		
		
		
		
		
		
	\end{thebibliography}
		\end {document}